\documentstyle[12pt,aasms4,flushrt,tighten]{article}

\newcommand{\be}{\begin{equation}}
\newcommand{\ee}{\end{equation}}
\newcommand{\ngc}{NGC~6090}

\newcommand{\kms}{km\,s$^{-1}$}

\newcommand{\ie}{{\it i.e.\/}}

\newcommand{\etal}{et~al.}
\newcommand{\hst}{{\it HST\/}}

\begin{document}

\bigskip
\title{ NICMOS Observations of Interaction Triggered Star Formation in
        the Luminous Infrared Galaxy NGC 6090\altaffilmark{1}}

\author{Nadine Dinshaw,\altaffilmark{2} 
Aaron S. Evans,\altaffilmark{3}
Harland Epps,\altaffilmark{2}
Nick Z. Scoville,\altaffilmark{3}
\and
Marcia Rieke\altaffilmark{4}
}
\altaffiltext{1}{Based on observations with the NASA/ESA {\it Hubble Space
        Telescope,} obtained at the Space Telescope Science Institute,
        which is operated by the Association of Universities for
        Research in Astronomy, Inc., under NASA contract NAS5-26555.}
\altaffiltext{2}{UCO/Lick Observatory, University of California, 
	Santa Cruz, CA 95064}
\altaffiltext{3}{California Institute of Technology, Pasadena, CA 91125}
\altaffiltext{4}{Steward Observatory, University of Arizona, Tucson, 
	AZ 85721}

\begin{abstract}

High resolution, 1.1, 1.6, and 2.2 $\mu$m imaging of the luminous infrared
galaxy \ngc\ obtained with NICMOS of the {\it Hubble Space Telescope\/}
are presented.  These new observations are centered on the two nuclei of
the merger, and reveal the spiral structure of the eastern galaxy and the
amorphous nature of the western galaxy. The nuclear separation of 3.2~kpc
($H_0 = 75$~\kms~Mpc$^{-1}$) indicates that \ngc\ is at an intermediate
stage of merging. Bright knots/clusters are also visible in the region
overlapping the merging galaxies; four of these knots appear bluer than
the underlying galaxies and have colors consistent with young ($\le
10^7$~yr) star clusters.  The spatial coincidence of the knots with the
molecular gas in NGC 6090 indicates that much of the present star
formation is occuring outside of the nuclear region of merging galaxies,
consistent with recent studies of other double nuclei luminous infrared
galaxies.

\end{abstract}


\section{Introduction}
\label{sect-intro}

Luminous galaxies have infrared luminosities $L_{\rm IR} (8-1000\mu{\rm
m}) \gtrsim 10^{11} L_\odot$, making them some of the most luminous
objects in the local universe.  Interactions and mergers of gas-rich
spirals are thought to play a significant role in production of the high
infrared luminosities, although the exact nature of the luminosity source
is still the subject of some controversy. Two of the more popular
candidates for the energy source are dust-enshrouded active galactic
nuclei (AGN) or circumnuclear starbursts (Rieke \etal\ 1985; Joseph \&
Wright 1985; Sanders \etal\ 1988; Sanders, Scoville \& Soifer 1991), both
of which are believed to be fuelled by the large quantities of molecular
gas concentrated at the merger nucleus.  It is likely that both mechanisms
play some role, and recent studies have focused on which of the two is the
dominant mechanism (Sanders \& Mirabel 1996 and references therein, Smith
\etal\ 1998; Genzel \etal\ 1998; Lutz et al. 1998).

The luminous infrared galaxy NGC 6090 (Mrk~496 = UGC~10267; $L_{\rm IR}
\simeq 3 \times 10^{11} L_\odot$: e.g. Acosta-Pulido et al. 1996) is an
interacting system in an intermediate stage of merging. At optical
wavelengths, \ngc\ appears as a double nucleus system. The separation of
the radio nuclei is 5\farcs4, corresponding to a projected distance of
3.2~kpc. There is considerable evidence for starburst activity, but no
evidence at optical and radio wavelengths for a compact AGN. Calzetti \&
Kinney (1992) used the dereddened H$\alpha$ emission line and $L_{\rm B}$
to estimate the star formation rates in \ngc\ and show that the
interacting system is undergoing a strong burst of star formation. Based
on the presence of deep 2.3\,\micron\ CO absorption, Ridgway \etal\ (1994)
argued that the near-infrared emission is probably dominated by late-type
supergiants or metal-rich giant stars, as opposed to an AGN. Finally,
Batuski \etal\ (1992) found the 6~cm radio emission is spatially extended
($\sim$\,3.9~kpc in extent), more indicative of a starburst origin than
AGN. The emission line spectra of both nuclei have been classified as {\sc
H~ii} based on the ratios of the [{\sc N~ii}], [{\sc S~ii}] and [{\sc
O~i}] to H$\alpha$ lines (Mazzarella \& Boroson 1993; see also, Veilleux
\etal\ 1995), consistent with the starburst classification by
Acosta-Pulido et al. (1996) based on the mid-infrared PAH features.

\ngc\ is one of a large sample of luminous infrared galaxies to be studied
as part of a program to investigate their morphologies and the source of
their extremely high infrared luminosity (Scoville et al. 1999). In this
paper, we concentrate on high resolution imaging of \ngc\ in three filters
centered at 1.1\,\micron, 1.6\,\micron\ and 2.2\,\micron\ obtained using
the Near Infrared Camera and Multiobject Spectrometer (NICMOS) on the {\it
Hubble Space Telescope\/}.  These images reveal new features not seen in
the ground-based optical and infrared data, such as bright knots of star
formation and a stellar bar in the primary galaxy. These features are
compared with similar evidence of star formation in other double nuclei
luminous infrared galaxies. Throughout this paper, we use $H_0 =
75$~\kms~Mpc$^{-1}$ and adopt a distance of 122.0~Mpc for \ngc\ (Condon
\etal\ 1990).

\section{Observations and Reductions}
\label{sect-obs}

\ngc\ was observed with camera 2 of NICMOS onboard the {\it Hubble
Space Telescope\/} on 10 November 1997 UT. Camera 2 is a $256\times
256$ HgCdTe array. It has a plate scale of $0\farcs0762\ \rm
pixel^{-1}$ in $x$ and $0\farcs0755\ \rm pixel^{-1}$ in $y$, resulting
a field of view of $19\farcs5 \times 19\farcs3$ (Thompson \etal\
1998). Exposures of \ngc\ were taken using the F110W, F160W and F222M
filters, centered at 1.1\,\micron, 1.6\,\micron\ and 2.22\,\micron,
respectively. The data are diffraction limited with resolutions of
0\farcs11 (F110W), 0\farcs16 (F160W) and 0\farcs22 (F222M).  The data
were taken in {\tt MULTIACCUM} mode where the detector is read out
nondestructively at intermediate stages of the integration, making it
easier to identify and remove cosmic ray spikes during the reduction
phase. The data were obtained using a four-point spiral dither
pattern, with a 1\farcs9125 (25.5 pixel) dither step. The exposure
time at each dither position was 96 seconds in the F110W and F160W
filters and 136 seconds in the F222M filter.  Similarly, F222M images
offset from \ngc\ by 95\arcsec\ were obtained during the same orbit
and immediately after the target observations in a three-point spiral
dither to measure the thermal background. The thermal background in
F222M amounted to 1.8~ADU~s$^{-1}$, and was negligible in F110W and
F160W.

Data reduction was carried out using the NICMOS calibration pipeline
routine CALNICA developed by H.\ Bushouse at STScI (see Bushouse
1997). The reductions included dark current subtraction, non-linearity
correction and flat-fielding using on-orbit flats. Refined offsets
between the individual images for a given filter were obtained using
the pipeline routine CALNICB. The images were combined using
Variable-Pixel Linear Reconstruction (informally known as
``drizzling") developed for undersampled, dithered images by A.\
Fruchter and R.\ Hook.  The algorithm first shrinks the input pixels
before projecting them onto a finer output grid (see Hook \& Fruchter
1997 for a more detailed description of the technique). The pixels
were oversampled by a factor of 2 which resulted in a halving of the
pixel scale of the final combined image.

Flux calibration was achieved using the conversion factors $2.03 \times
10^{-6}$, $2.19 \times 10^{-6}$ and $5.49 \times 10^{-6}$ Jy/ADU/s for
F110W, F160W and F222M, respectively, and corresponding magnitude
zero-points of 1775, 1083 and 668~Jy (Rieke et al. 1999). The calibration
yields a photometric accuracy of less than 10\%. The measured
fluxes in the F160W and F222M filters (which can be regarded as $H$ and
$K$, respectively) for a 10\arcsec\ diameter beam centered on the peak of
the primary galaxy agree to within 10\% with ground-based observations of
\ngc\ by Carico \etal\ (1990). Interpolating the 1.1\,\micron\ and
1.6\micron\ fluxes to get a measure of the $J$ magnitude, we find a
similar agreement.

Observations of a point spread function (PSF) star were not obtained
for \ngc\ because all the nearby guide stars resided outside a
2\arcsec\ radius of \ngc, which would have required a target
re-acquisition with an associated time penalty we could not afford.
However, observations of a PSF star, obtained for another galaxy,
exist that were taken in an identical manner as the \ngc\ data and in
the same filters. Those data were obtained one day after the \ngc\
observations (11 November 1997 UT).

The images are diffraction limited, so the resolutions in the different
filters are not the same. For this reason, we deconvolved the images
beyond the resolution of the 1.1\,\micron\ image using the Richardson-Lucy
algorithm and the observed PSF star.  Figure 1 shows a three-color
composite of the deconvolved images which have been resmoothed to the same
resolution (0\farcs13) using a Gaussian transfer function. The
Richardson-Lucy algorithm does not conserve flux, especially in the
presence of a background, therefore, we have not used the deconvolved
images for any of the flux measurements made in this paper.  Gray-scale
representations of the undeconvolved images and their respective contour
diagrams are shown in Figure 2.

\section{Results}
\label{sect-results}

\subsection{Morphology and Nuclear Structure}
\label{sect-morph}

Figures 1 and 2 clearly show that NGC 6090 is in an intermediate stage of
interaction. The galaxies are widely separated (3.2~kpc) compared to the
separation of the nuclei of Arp~220 ($\sim$\,350~pc) which is thought to
be in an advanced stage of merger. The optical image of \ngc\ in
Mazzarella \& Boroson (1993) shows two galaxies of roughly equal size that
are almost completely overlapping and a pair of tidal tails with a full
extent of $\approx$ 50 kpc (80$\arcsec$).  The NICMOS images are centered
on the nuclei - while the southwestern nucleus  (\ngc W) appears
amorphous, the northeastern galaxy (\ngc E) clearly has distorted spiral
structure.  The most striking feature of \ngc E is the abundance of
luminous blue and red knots along the western arm (i.e., the side of the
galaxy closest to \ngc W) and the paucity of similar clusters on its
eastern spiral arms. \ngc W also shows a similar abundance of blue knots
along the western side, and one extremely luminous knot at the northern
end. For simplicity, we will refer to the regions containing the knots as
the overlap region.

\subsection{Registration}
\label{sect-reg}

Figure 3 shows contours of the 1.49~GHz radio continuum
emission overlaid on the 2.22\micron\ image. The radio data with a
resolution of 1\farcs5 were published in Condon et al.\ (1990) and
obtained via the NASA/IPAC Extragalactic Database.
\footnote{The NASA/IPAC Extragalactic Database (NED) is operated by
     the Jet Propulsion Laboratory, California Institute of
     Technology, under contract with the National Aeronautics and
     Space Administration.}
The astrometric precision of \hst\ is insufficient to register the radio
contours with the NICMOS image using their absolute positions.  Therefore,
radio coordinates were compared with astrometry of the NGC 6090 nuclei
derived from the positions of USNO-A1.0 database
stars within a 7$\arcmin\times7\arcmin$ {\it R}-band image of the merger
(Mazzarella, private comm.).  Condon \etal\ (1990) and Hummel et al.
(1987) measured the coordinates of the northeastern radio peak to be
$\alpha_{1950} = 16^{\rm h}10^{\rm m}24\fs56$, $\delta_{1950} =
+52^\circ35^\prime05\farcs2$ and $\alpha_{1950} = 16^{\rm h}10^{\rm
m}24\fs54$, $\delta_{1950} = +52^\circ35^\prime05\farcs3$, respectively.
These coordinates are consistent with {\it R}-band
coordinates of the northeastern nucleus, i.e., $\alpha_{1950} = 16^{\rm
h}10^{\rm m}24\fs57$, $\delta_{1950} = +52^\circ35^\prime04\farcs8$.
However, it appears from our high-resolution NICMOS data that the
southwestern radio nucleus does not correspond to the bright point source
in the southwestern galaxy, thus this source is not likely the nucleus of
\ngc W. In fact, there is evidence for enhanced diffuse emission near the
southwestern radio nucleus that is a more likely candidate for the nuclear
region of \ngc W. This enhanced emission is more apparent in the
2.2\,\micron\ contour plot in Figure 2.

\subsection{Nuclear and Global Infrared Colors}
\label{sect-red}

Figure 4 shows ($\rm m_{1.1} - m_{1.6}$) and ($\rm m_{1.6} - m_{2.2}$)
color maps of \ngc\ constructed from the ratios of the 1.1\,\micron,
1.6\,\micron, 2.22\,\micron\ images. In order to avoid spurious color
variations due to the mismatch in the PSFs among the different filters,
the shorter wavelength image was convolved with a Gaussian so that the
PSFs at both wavelengths matched before taking the ratio. Some residual
structure from the first diffraction ring is still apparent in Figure 4
for the bright point source in \ngc W. In order to minimize amplification
of the noise in the color maps, only regions with signal greater than six
times the noise are displayed.

The ($\rm m_{1.1} - m_{1.6}$) color ranges from 1.0~mag in the nuclear
region of \ngc E to 0.7~mag in the outer spiral arms which are defined
by prominent dust lanes. Aside from the spiral pattern, \ngc E has a
relatively smooth and flat color structure that suggests that dust, if
present, is uniformly distributed across the galaxy.  The ($\rm
m_{1.6} - m_{2.2}$) color is similarly flat, roughly $0.2 - 0.3$~mag
across the galaxy. In contrast, \ngc W shows small-scale variations in
color that are probably due to the presence of faint knots, but
neither color map shows obvious organized structure. Its average
colors are ($\rm m_{1.1} - m_{1.6}) = 0.7 - 0.8$~mag and ($\rm m_{1.6}
- m_{2.2}) = 0.1 - 0.2$~mag.  The colors derived for \ngc W and E are
consistent with those typical of starburst galaxies and a normal
late-type stellar population, $J-H = 0.6 - 0.8$~mag and $H-K = 0.1 -
0.4$~mag (Hunt \etal\ 1997). Ridgway \etal\ (1994) have detected a
strong 2.3\,\micron\ CO absorption feature in \ngc. Since this feature
is produced in the atmospheres of red giants and supergiants, the near
infrared emission is most likely dominated by supergiant or metal-rich
giant stars with some reddening by interstellar dust (Ridgway \etal\
1994). 

\subsection{Cluster Colors}
\label{sect-knots}

As mentioned in \S3.1, the new high-resolution images of \ngc\ reveal a
number of luminous knots, or clusters, several of which appear to be much
bluer ($\rm m_{1.1} - m_{1.6} < 0.6$) than the underlying galaxies.  The
positions, apparent magnitudes and colors of 12 knots measured in \ngc W
and E are given in Table~1. The positions are listed as offsets from the
peak emission in the 2.22\micron\ image.  Table~1 only includes those
knots with magnitudes brighter than 20~mag for which we were able to
obtain reliable flux measurements. The magnitudes were calculated from
aperture photometry using the IRAF task APPHOT.  Because of potential
confusion from the background, the flux in each knot was first measured in
a small aperture (typically $4 - 6$ pixels in radius) and adjusted later
based on aperture corrections tables computed from the observed PSF star.
The background galaxy flux was estimated using the median of the flux in a
$3 - 5$-pixel annulus located at a radius of typically $7 - 11$ pixels
from the center of the knot.  This was far enough from the source to
ensure that the flux in the far wings of the diffraction pattern was not
included in the background measurement, but close enough to obtain a
measure of the true surrounding flux.

In order to estimate the uncertainty in the photometry, we simulated
the data by scaling the PSF star to magnitudes ranging from the
brightest ($\sim 16$ mag) to the faintest ($\sim 20$ mag) observed
clusters and adding the scaled PSFs into various regions in the data
of \ngc. We then tried to recover the known flux using our aperture
photometry technique. Not surprisingly, the largest uncertainty at
faint magnitudes is caused by difficulty in determining the background
level. At 18 mag and brighter, we found the uncertainty to be 0.1~mag
in all three bands, consistent with the formal error in the flux due
to poisson noise and detector readout. This is also consistent with
the uncertainties in the photometric calibrations. For magnitudes
between 18 and 20~mag, the uncertainty grows to as much as 0.30~mag at
1.1 and 1.6\,\micron. The error is slightly worse at 2.22\,\micron\
because of a combination of poorer signal-to-noise data and broader
PSF. Since we are interested in the colors of the clusters, we have
tried to be consistent in our measurements of the cluster fluxes among
the different filters. Given that we expect the errors to be
systematic, we estimate the associated uncertainty in the colors to be
roughly the same as that of the flux measurements, \ie\ $0.1 -
0.3$~mag.

The average colors of clusters 1 to 8 in Table~1 are $J-H = 0.60$ and $H-K
= 0.17$, consistent with the colors of an old stellar population.  The
colors are similar to those of the old globular cluster systems around the
galaxy NGC~5128 with ages $> 10^9$~yr (Frogel 1984), but are also
consistent with the colors of highly reddened young star clusters. In
addition to the red clusters, there are 4 relatively blue clusters in the
spiral arm of \ngc E which have colors consistent with a younger
population of stars.  The average $J-H$ color for clusters 9 to 12 is
$\sim$\,0.18~mag.  The average $H-K$ color is $0.43$~mag, but it has a
large spread probably reflecting the fact that the uncertainties can be
quite large in $K$ at the faintest magnitudes.  Compared to the typical
colors for a population of stars with ages $\le 6\times 10^6$~yr, $J-H
\simeq 0.25$ and $H-K \simeq 0.50$~mag, the clusters appear to be
consistent with young star-forming regions with little infrared reddening
by interstellar dust. Two colors are not sufficient to fully constrain the
ages of the clusters, and spectroscopy is needed to confirm whether they
are really young stars reddened by interstellar dust or evolved unreddened
stars.

The brightest point source in \ngc W is considerably brighter than any
of the other clusters in the galaxy. In fact, in the 1.1\,\micron\ and
1.6\,\micron\ images, it has a peak brightness that is greater than
that of the nucleus of \ngc E. We have assumed throughout this section
that this source is a very luminous star cluster. This is reasonable
assumption based on its colors, which are consistent with the rest of
the clusters in \ngc W, but we cannot rule out that it may be a
foreground star.

\section{Discussion}

The close nuclear separation, the high infrared luminosity, the extended
optical tidal tails and the double nuclei of NGC 6090 are all evidence of
galaxy-galaxy interaction. Of the two merging systems, \ngc E most
resembles its progenitor -  a spiral disk seen face-on - with evidence for
a stellar bar in its inner disk.  The bar is most evident in the
2.22\,\micron\ contour plot in Figure 2.  Numerical simulations show that
during a merger, tidal forces from a companion galaxy trigger the
formation of a bar in the disk of the perturbed galaxy (Noguchi 1988;
Shlosman, Frank \& Begelman 1989, Shlosman, Begelman \& Frank 1990). The
bar acts to trigger starburst activity by rapidly funneling large amounts
of gas to the nuclear regions.  A correlation between stellar bars and
stellar activity appears to be present for Seyfert galaxies (Maiolino
\etal\ 1997), and there is evidence that a significant fraction of
starburst galaxies may contain stellar bars in their disks (Colina
\etal\ 1997, Contini, Considere \& Davoust 1998). The star formation rate
derived for \ngc E by Calzetti \& Kinney (1992) indicates the galaxy is
undergoing an intense burst of star formation, where the present star
formation rate (SFR) is a factor of ten above the average SFR. If the SFR
is primarily associated with the nucleus of NGC 6090E, the nuclear
starburst region is only 135~pc ($\sim$\,0\farcs25) in extent.

The evidence for luminous clusters in the overlap region of NGC 6090
indicates that most of the star formation visible at these wavelengths may
actually be off-nuclear . This is supported by
millimeter (CO) observations of this merger, which show a single dominant
component of molecular gas, centered approximately midway between the two
nuclei (Bryant \& Scoville 1999).  The CO emission appears elongated and
aligned along the direction between the radio nuclei (P.A. = 60$^\circ$).
Using the standard conversion factor, the molecular gas mass is determined
to be $1.4 \times 10^{10}$~M$_\odot$ (Sanders, Scoville \& Soifer 1991).
In contrast, the velocity width of CO emission is narrow ($\Delta v =
136$~\kms\ FWHM), indicating dynamical mass of $M_{\rm dyn} = 4.60 \times
10^9$~M$_\odot$ for the molecular gas complex, although it is unlikely to
be self-gravitating or virialized (Bryant 1996).  Given the abundance of
molecular gas and its spatial coincidence with the putative young star clusters,
the most likely scenario for the activity in the overlap region is that
the gas has been stripped from the progenitors, and that
localized star formation has resulted (e.g. Barnes \& Hernquist 1996; Mihos \&
Hernquist 1996).

The presence of molecular gas and star formation in the overlap
region of merging galaxies may be quite common. Both the luminous infrared
galaxies NGC 6240 (Bryant \& Scoville 1999; Tacconi et al. 1999) and VV114
(Yun, Scoville, \& Knop 1994; Frayer et al.  1999) have most of their
molecular gas and dust mass in the overlap region. Recent NICMOS images of
VV114 show the presence of numerous star clusters between the two nuclei
(Scoville et al. 1999). The slightly less luminous merger Arp 244 (aka:
the Antennae galaxies) also has a substantial fraction of its molecular
gas concentrated between the two nuclei (Stanford et al. 1990), and recent
ISO imaging and spectroscopy that shown that the most significant
starburst occuring in that merger is occuring the the overlap region
(Mirabel et al. 1998). However, this morphology is not ubiquitous among
mergers which still harbor separated nuclei - the ultraluminous ($L_{\rm
IR} > 10^{12}$ L$_\odot$) infrared galaxy Arp 220, which is a
comparatively advanced merger, has a significant amount of its molecular
gas and mid-infrared emission associated with the nuclei (Sakamoto et al.
1999; Soifer et al.  1999). Thus, the presence of star formation in the
overlap region likely depends on the structure of the merging galaxies and
the stage of the merger (e.g.  Mihos \& Hernquist 1996).

\section{Summary} \label{sect-discuss}

We have obtained high resolution images of the luminous infrared galaxy
\ngc\ with NICMOS of the {\it Hubble Space Telescope\/} in three broadband
filters centered at 1.1, 1.6 and 2.22\,\micron. The images show the
eastern galaxy to have spiral structure and a nuclear bar, and the western
galaxy to be amorphous.  The galaxies involved in the interaction exhibit
bright blue knots which we interpret as star forming knots. The spatial
coincidence of the knots and the molecular gas in the region overlapping
the two merging galaxies is similar to that observed in other luminous
infrared galaxies, and provides evidence that much of present star
formation in intermediate stage mergers may occur outside of the nuclei of
mergers.

\acknowledgments The NICMOS project is supported by NASA grant NAG 5-3042.
We are grateful to Lee Armus, Chris Fryer, Chris Mihos and Brian Sutin for
useful conversations. We thank Dave Sanders, D.-C. Kim, and J. Mazzarella
for making their ground-based $K$ image of \ngc\ available to us. Finally,
we thank J. Mazzarella for providing astrometry of NGC 6090.

\pagestyle{empty}

\begin{deluxetable}{lrrcccccc}
\tablewidth{0pc}
\tablecaption{Knots in NGC~6090}
\tablehead{
\colhead{ No. }                 &
\colhead{ $\Delta \alpha^a$ }   &
\colhead{ $\Delta \delta^a$ }   &
\colhead{ m$_{1.1}^b$ }         &
\colhead{ m$_{1.25}^c$ }        &
\colhead{ m$_{1.6}^b$ }         &
\colhead{ m$_{2.2}^b$ }         &
\colhead{ m$_{1.25-1.6}$  }     &
\colhead{ m$_{1.6-2.2}$  }      }
\startdata
  1 & W 5.78 & $-$3.09 & 17.23 & 17.00 & 16.39 & 15.87 & 0.69 & 0.23 \nl
  2 & W 5.49 & $-$3.84 & 20.11 & 20.00 & 19.63 & 19.25 & 0.45 & 0.10 \nl
  3 & W 5.02 & $-$4.72 & 19.69 & 19.45 & 18.84 & 18.50 & 0.69 & 0.06 \nl
  4 & W 4.92 & $-$5.03 & 19.93 & 19.73 & 19.18 &  ---  & 0.63 & ---  \nl
  5 & W 5.17 & $-$3.57 & 19.81 & 19.60 & 19.03 & 18.79 & 0.65 & -0.04 \nl
  6 & W 5.21 & $-$3.28 & 19.70 & 19.53 & 19.04 & 18.56 & 0.57 & 0.20 \nl
  7 & W 2.56 &    1.70 & 19.79 & 19.68 & 19.31 &  ---  & 0.45 & ---  \nl
  8 & W 1.84 &    1.98 & 19.23 & 19.01 & 18.42 & 17.67 & 0.67 & 0.47 \nl
  9 & W 0.96 &    0.76 & 19.13 & 19.11 & 19.03 & 18.55 & 0.16 & 0.10 \nl
 10 & W 1.16 &    0.34 & 18.73 & 18.71 & 18.59 & 18.08 & 0.20 & 0.13 \nl
 11 & W 1.62 & $-$0.40 & 18.24 & 18.25 & 18.18 & 17.60 & 0.15 & 0.30 \nl
 12 & W 0.78 & $-$1.42 & 20.14 & 20.13 & 20.00 & 18.55 & 0.21 & 1.17 \nl
\enddata
\hskip -6mm
\tablenotetext{a}{Offsets are in arcseconds relative
to the peak of NGC\,6090E in the 2.2\micron\ image,
$\alpha_{1950} =  16^{\rm h}10^{\rm m}24\fs56$,
$\delta_{1950} = +52^\circ35^\prime05\farcs2$.
\vskip 1mm}
\tablenotetext{b}{Magnitudes have typical uncertainties of $\pm0.1-0.3$~mag.
\vskip 1mm }
\tablenotetext{c}{Magnitudes interpolated from m$_{1.1}$ and
m$_{1.6}$ magnitudes.}
\end{deluxetable}

\begin{figure}[p]
\caption[]{ Three-color composite image of the 1.1, 1.6 and
2.22\micron\ data for a 15\arcsec\ field of view centered on \ngc. The
intensity scale is logarithmic. The images have been deconvolved using
the Richardson-Lucy algorithm beyond the resolution of the 1.1\micron\
data and then smoothed with a Gaussian transfer function to a
resolution of 0\farcs13. North is up; east is to the left.}
\label{1}
\end{figure}

\begin{figure}[p]
\caption[]{ Gray-scale representations of the individual 1.1, 1.6 and
2.22\micron\ images and their corresponding contour diagrams. The
images are plotted with their original resolutions: 0\farcs11 (F110W),
0\farcs16 (F160W) and 0\farcs22 (F222M). The plot axes are labeled in
arcseconds offset from the peak of \ngc E in the 2.2\micron\ image,
$\alpha_{1950} = 16^{\rm h}10^{\rm m}24\fs56$, $\delta_{1950} =
+52^\circ35^\prime05\farcs2$. The contours are spaced logarithmically
by factors of 1.69, 1.65 and 1.38 starting at the 0.11, 0.11 and
0.37~$\mu$Jy levels at 1.1, 1.6 and 2.22\,\micron.}
\label{fig-gray}
\end{figure}


\begin{figure}[p]
\caption[]{ Contours of 1.49 GHz radio continuum emission superimposed
on the 2.22\micron\ image. The radio contours are from Condon et al.\
(1990). The registration of the radio and infrared data are discussed
in the text. The plot axes are labeled in arcseconds offset from the
peak of \ngc E in the 2.2\micron\ image, $\alpha_{1950} = 16^{\rm
h}10^{\rm m}24\fs56$, $\delta_{1950} = +52^\circ35^\prime05\farcs2$.
}
\label{fig-radio}
\end{figure}

\begin{figure}[p]
\caption[]{ Color maps of \ngc\ in ($\rm m_{1.1} - m_{1.6}$) and ($\rm
m_{1.6} - m_{2.22}$) constructed from the ratios of the 1.1\,\micron,
1.6\,\micron, 2.22\,\micron\ images. Colors range from 0.30 to
1.25~mag (blue to red) in ($\rm m_{1.1} - m_{1.6}$) and $-$0.28 to
1.14~mag in ($\rm m_{1.6} - m_{2.22}$). 
}
\label{fig-contour}
\end{figure}








\newpage
\begin{references}

\reference{} Acosta-Pulido, J. A. et al. 1996, A\&A, 315, L121

\reference{} Barnes, J., \& Hernquist, L. 1992, \araa, 30, 705

\reference{} Barnes, J., \& Hernquist, L. 1996, \apj, 471, 115

\reference{} Batuski, D. J., Hanisch, R. J., \& Burns, J. O. 1992, \aj,
103, 1077

\reference{} Bruzual, A. G., \& Charlot, S. 1993, \apj, 405, 538

\reference{} Bryant, P. M. \& Scoville, N. Z. 1999, ApJ, in press

\reference{} Bushouse, H. 1997, in HST Calibration Workshop,
eds. S. Casertano, R. Jedrzejewski, T. Keyes, \& M. Stevens
(Baltimore), 223

\reference{} Calzetti, D., \& Kinney, A. L. 1992, \apj, 399, L39

\reference{} Carico, D. P., Sanders, D. B., Soifer, B. T., Elias,
J. H., Matthews, K., \& Neugerbauer, G. 1990, \aj, 95, 356

\reference{} Colina, L., Gar\'{i}a Vargas, M. L., Mas-Hesse, J. M.,
Alberdi, A., \& Krabbe, A. 1997, \apj, 484, L41

\reference{} Condon, J. J., Helou, G., Sanders, D. B., \& Soifer,
B. T. 1990, \apjs, 73, 359

\reference{} Contini, T., Considere, S., \& Davoust, E. 1998, \aap,
130, 285

\reference{} Genzel, R., Lutz, D., Sturm, E., Egami, E., Kunze, D.,
Moorwood, A. F. M., Rigopoulou, D., Spoon, H. W. W., Sternberg, A.,
Tacconi-Garman, L. E., Tacconi, L., \& Thatte, N. 1998, \apj, 498, 579

\reference{} Helou, G., Soifer, B. T., \& Rowan-Robinson, M. 1985,
\apj, 298, L7

\reference{} Hook, R. N., \& Fruchter, A. S. 1997, Astronomical Data
Analysis Software and Systems VI, A.S.P. Conference Series, (Hunt,
G. \& Payne, H. E., eds.), 125, p 147

\reference{} Hummel, E., van der Hulst, J. M., Keel, W. C.,
Kennicutt, Jr., R. C. 1987, A\&AS, 70, 517

\reference{} Hunt, L. K., Malkan, M. A., Salvati, M., Mandolesi, N.,
Palazzi, E., \& Wade, R. 1997, \apjs, 108, 229

\reference{} Joseph, R. D. \& Wright, G. S. 1985, \mnras, 214, 87

\reference{} Klaas, U., Haas, M., Heinrichsen, I., \& Schulz, B.,
A\&A, 325, L21

\reference{} Maiolino, R., Ruiz, M., Rieke, G. H., \& Papadopoulous,
P. 1997, \apj, 485, 552

\reference{} Mazzarella, J. M., \& Boroson, T. A. 1993, \apjs, 85, 27

\reference{} Mirabel, I. F., et al. 1998, A\&A, 333, L1

\reference{} Mihos, J. C. \& Hernquist, L. 1996, ApJ, 464, 641

\reference{} Noguchi, M. 1991, \mnras, 251, 360

\reference{} Ridgway, S. E., Wynn-Williams, C. G., \& Becklin,
E. E. 1994, \apj, 428, 609

\reference{} Lutz, D., Spoon, H. W. W., Rigopoulou, D., Moorwood, A. F. M.,
\& Genzel, R. 1998, \apj, 505, 103

\reference{} Rieke, G. H., Cutri, R. M., Black, J. Hl, Kailey, W. F.,
McAlary, C. W., Lebofsky, M. J., \& Elston, R. 1985, \apj, 290, 116

\reference{} Rieke, M. et al. 1999, in prep.

\reference{} Sakamoto, K., Scoville, N. Z., Yun, M. S., Crosas, M.,
Genzel, R., \& Tacconi, L. J. 1999, ApJ, 514, 68

\reference{} Sanders, D. B., \& Mirabel, I. F. 1996, \araa, 34, 749

\reference{} Sanders, D. B., Scoville, N. Z., \& Soifer, B. T. 1991,
\apj, 370, 158

\reference{} Sanders, D. B., Soifer, B. T., Elias, J. H., Madore,
B. F., Matthews, K., Neugebauer, G., \& Scoville, N. Z. 1988, \apj,
325, 74

\reference{} Scoville, N. Z. et al. 1999, submitted

\reference{} Shlosman, I., Begelman, M. C., \& Frank, J. 1990, Nature,
345, 679

\reference{} Shlosman, I., Frank, J., \& Begelman, M. C. 1989, Nature,
338, 45

\reference{} Smith, H. E., Lonsdale, C. J., Lonsdale, C. J., \&
Diamond, P. J. 1998, \apj, 493, L17

\reference{}  Soifer, B. T., Neugebauer, G., Matthews, K., Becklin, E. E.,
Ressler, M., Werner, M. W., Weinberger, A. J., \& Egami, E.  1999,
ApJ, 513, 207

\reference{} Standford, S. A., Sargent, A. I., Sanders, D. B., \& Scoville,
N. Z. 1990, ApJ, 349, 492
 
\reference{} Tacconi, L. J., Genzel, R., Tecza, M., Gallimore, J. F.,
Downes, D., \& Scoville, N. Z. 1999, ApJ, in press

\reference{} Thompson, R. I., Rieke, M., Schneider, G., Hines, D. C., \&
Corbin, M. R. 1998, ApJ, 492, L95

\reference{} Veilleux, S., Kim D.-C., Sanders, D. B., Mazzarella,
J. M., \& Soifer, B. T. 1995, \apjs, 98, 171

\reference{} Yun, M. S., Scoville, N. Z., \& Knop, R. A. 1994, ApJ, 430, L109
 
\end{references}
\end{document}